\documentclass[aps,pre,twocolumn,groupedaddress]{revtex4-1}

\usepackage{amsfonts,amssymb,amsmath} 
\usepackage{color} 
\usepackage{graphicx} 
\usepackage{epstopdf,mathrsfs} 
\usepackage[colorlinks,linktocpage,linkcolor=blue]{hyperref}

\begin{document}


\title{Langevin dynamics for L\'{e}vy walk with memory}
\author{Yao Chen}
\author{Xudong Wang}
\author{Weihua Deng}

\affiliation{School of Mathematics and Statistics, Gansu Key Laboratory
of Applied Mathematics and Complex Systems, Lanzhou University, Lanzhou 730000,
P.R. China}



\begin{abstract}
Memory effects, sometimes, can not be neglected. In the framework of continuous time random walk, memory effect is modeled by the correlated waiting times. In this paper, we derive the two-point probability distribution of the stochastic process with correlated increments as well as the one of its inverse process, and present the Langevin description of L\'{e}vy walk with memory, i.e., correlated waiting times. Based on the built Langevin picture, the properties of  aging and nonstationary are discussed. The Langevin system exhibits sub-ballistic superdiffusion if the friction force is involved, while it displays super-ballistic diffusion or hyperdiffusion if there is no friction. It is discovered that the correlation of waiting times suppresses the diffusion behavior whether there is friction or not, and the stronger the correlation of waiting times becomes, the slower the diffusion is. In particular, the correlation function, correlation coefficient, ergodicity, and scaling property of the corresponding stochastic process are also investigated.


\end{abstract}


\maketitle

\section{Introduction}
In recent decades, it has been widely discovered that anomalous diffusions are ubiquitous especially in complex environment. Anomalous diffusion is in general characterized by the nonlinear evolution in time of the mean squared displacement (MSD) of a particle, i.e.,
\begin{equation}
  \langle (\Delta x(t))^2\rangle=\langle[x(t)-\langle x(t)\rangle]^2\rangle\simeq t^\alpha  \quad  (\alpha\neq1),
\end{equation}
which represents subdiffusion for $0<\alpha<1$ and superdiffusion for $\alpha>1$; for the case $\alpha=2$, it is called ballistic diffusion, and super-ballistic diffusion is for $\alpha>2$ (or hyperdiffusion \cite{MetzlerKlafter:2000,BouchaudGeorges:1990}).

One of the powerful and popular models to describe anomalous diffusion is the continuous time random walk (CTRW), which was originally introduced by Montroll and Weiss in 1965 \cite{MontrollWeiss:1965}, extending regular random walks on lattices to a continuous time variable. It has been successfully applied in various fields, including the charge carrier transport in amorphous semiconductors \cite{ScherMontroll:1975}, electron transfer \cite{Nelson:1999}, dispersion in turbulent systems \cite{SolomonWeeksSwinney:1993}, and so on.
CTRWs contain two types of independent identically distributed (i.i.d.) random variables: jump lengths and waiting times. If both the second moment of jump lengths and the first moment of waiting times are finite, the scaling limit of the CTRW model leads to Brownian motion. On the other hand, if the second moment of jump lengths and/or the first moment of waiting times diverge(s),  we arrive at anomalous diffusion.

One of the representative models of anomalous diffusion is  L\'{e}vy flight \cite{ChechkinGoncharKlafterMetzlerTanatarov:2004,BrockmannGeisel:2003,DybiecGudowska:2009}, unfortunately which is less physical in the sense that a particle should have finite mass and velocity. 
As an amendment to it, L\'{e}vy walk \cite{ZaburdaevDenisovKlafter:2015,SanchoLacastaLindenbergSokolovRomero:2004,RebenshtokDenisovHanggiBarkai:2014,ZaburdaevDenisovHanggi:2013} with finite velocity seems to be more reasonable and originally characterized by coupled CTRWs \cite{ShlesingerKlafterWong:1982,KlafterBlumenShlesinger:1987,Zaburdaev:2006,ZaburdaevDenisovKlafter:2015}, in which the probability density functions (PDFs) of jump length and waiting time are coupled through a constant velocity. There are many applications of L\'{e}vy walk, for instance, anomalous superdiffusion of cold atoms in optical lattices \cite{KesslerBarkai:2012}, endosomal active transport within living cells \cite{ChenWangGranick:2015}, etc.

Another method to model the complex dynamics is using Langevin system. In 1994, Fogedby \cite{Fogedby:1994} introduced an equivalent form of the CTRW with power-law distributed waiting times, i.e., a Langevin equation coupled with a subordinator.
Since then, the concept of subordination \cite{Applebaum:2009}, which was put forward by Bochner \cite{Bochner:1949} in 1949, has gotten further developments. Especially in recent years, the coupled Langevin equations have been investigated in many literatures \cite{MeerschaertScheffler:2004,GajdaWylomanska:2015,WylomanskaKumarPoloczanskiVellaisamy:2016} as a useful tool to describe time-changed stochastic processes, which occur in many fields, such as biology \cite{GoldingCox:2006}, physics \cite{NezhadhaghighiRajabpourRouhani:2011}, ecology \cite{ScherMargolinMetzlerKlafterBerkowitz:2002}.

In many real life processes, the memory effects may and should be considered, e.g., the financial market dynamics \cite{Scalas:2006}, the bacterial motion \cite{RogersWalleWaigh:2008}, the movement ecology of animal motion \cite{Nathan_etal:2008}, etc. Then, the compound renewal processes \cite{Cox:1962} described by (coupled) CTRW models can not be used to characterize these phenomena. Recently, the correlated CTRWs \cite{ChechkinHofmannSokolov:2009,TejedorMetzler:2010,MagdziarzMetzlerSzczotkaZebrowski:2012_1,SchulzChechkinMetzler:2013}, where the jump length and/or waiting time are/is dependent on the preceding steps, are introduced and become a useful tool to describe the stochastic processes possessing memory effect. The equivalent Langevin picture for the limit process of the correlated CTRWs has been investigated in \cite{MagdziarzMetzlerSzczotkaZebrowski:2012_2}.

In this paper, we extend the correlated CTRWs in \cite{ChechkinHofmannSokolov:2009} to the L\'{e}vy-walk-type model with correlated waiting times. Based on the Langevin description, we fully discuss the diffusion behaviour and ergodicity of the stochastic process described by this model.
It has been showed in \cite{EuleZaburdaevFriedrichGeisel:2012} how the coupled Langevin equation describes ballistic diffusion exhibited by L\'{e}vy-walk-type model with independent power-law distributed waiting times.
Here, based on the underdamped Langevin system \cite{EuleZaburdaevFriedrichGeisel:2012}, by assuming the waiting times to be correlated, we get a sub-ballistic superdiffusion phenomenon; besides that, a hyperdiffusion phenomenon is observed in the non-friction case. The stronger the correlation of waiting times is, the slower the diffusion becomes whether there is friction or not. Note that the correlated waiting time process is no longer a subordinator due to the correlation, so the two-point probability distributions of this time process and its inverse process are not obvious. But according to the stationary and independent increments of L\'{e}vy process, the two-point probability distributions can be derived. Based on the two-point probability distributions of the correlated waiting time process and its inverse process, we calculate the correlation function of velocity process, and also discuss the correlation function, time averaged MSD of the corresponding stochastic process, finding the long-range dependence, weak ergodicity breaking, and scaling property. By making some comparisons between the L\'{e}vy-walk-type models with correlated and uncorrelated waiting times, we find that the correlation suppresses the diffusion behaviour and strengthens the ergodicity breaking.
Some simulations are presented to verify the theoretical results.


The structure of this paper is as follows. In Sec. \ref{two}, we review the one-point PDF of a stochastic process with correlated increments which can be regarded as the correlated waiting times in CTRW framework. Then we derive the two-point joint PDFs of this stochastic process together with its inverse process in Laplace space.
In Sec. \ref{three}, we propose the Langevin description of the L\'{e}vy-walk-type model with correlated waiting times. The model with strongly correlated waiting times is detailedly discussed through some physical observed quantities (such as, ensemble averaged MSD, correlation function, time averaged MSD) in Sec. \ref{IIIA} and Sec. \ref{IIIB}. And then the L\'{e}vy-walk-type model with weakly correlated waiting times is investigated in Sec. \ref{IIIC}.
Finally, we make some conclusions in section \ref{four}.

\section{Two-point joint probability distribution}\label{two}

At the beginning of this section, we briefly review the PDF of a stochastic process with correlated increments and that of its inverse process.
Such correlated increments have been investigated by Chechkin {\it et al.} \cite{ChechkinHofmannSokolov:2009} in CTRW framework as the correlated waiting times; the waiting time of the $n$-th step $\Delta t_n$ is defined as weighted sum of independent random variables:
\begin{equation}\label{1}
\Delta t_n=\sum_{j=0}^n M_{n,j}\tau_j.
\end{equation}
Here $M_{n,j}=M(n-j)$ is a memory function and the random variables $\{\tau_j\}$ are i.i.d. with
the characteristic function (i.e., the Laplace transform of PDF $p(\tau_j)$)
\begin{equation}
  p_\tau(\lambda):=\int_0^\infty d\tau_j e^{-\tau_j\lambda}p(\tau_j)=e^{-\lambda^\alpha}, \quad 0<\alpha<1.
\end{equation}
It is the existence of the weights $M_{n,j}$ in (\ref{1}), making $\Delta t_n$ to be correlated with each other. Hence, the specific form of $M_{n,j}$ determines the strength of correlation, while $M_{n,j}=\delta_{nj}$ corresponds to the weakest one, i.e., the uncorrelated case.
Equation (\ref{1}) can also be written into continuous form as
\begin{equation}\label{2}
\frac{dt(s)}{ds}=\int_0^sds'M(s-s')\tau(s'),
\end{equation}
where $M(s)$ is a non-negative continuous memory function and $\tau(s)$ is a white $\alpha$-stable L\'{e}vy noise with $0<\alpha<1$. 
The continuous form will be more convenient to be coupled into a Langevin system.
In the case of $M(s)=\delta(s)$, Eq. (\ref{2}) reduces to the uncorrelated waiting times obeying power-law distribution, i.e., $dt(s)/ds=\tau(s)$.
Applying the correlated waiting times, the following coupled Langevin equation can be introduced by modifying the uncorrelated case proposed by Fogedby \cite{Fogedby:1994} as
\begin{equation}
\frac{dx(s)}{ds}=\eta(s), \qquad
\frac{dt(s)}{ds}=\int_0^sds'M(s-s')\tau(s'),
\end{equation}
where $\eta(s)$ is a white Gaussian noise with mean value $\langle \eta(s)\rangle=0$ and $\langle \eta(s)\eta(s')\rangle=2D_v\delta(s-s')$. Assuming that $M(s)=s^{\beta-1}/\Gamma(\beta)$ is a power-law memory function with $0<\beta<1$, the MSD of the stochastic process described by the above coupled Langevin equation is $\langle x^2(t)\rangle\propto t^{\alpha/(\alpha\beta+1)}$ \cite{ChechkinHofmannSokolov:2009}, which is slower than $\langle x^2(t)\rangle\propto t^\alpha$, i.e., the case of uncorrelated waiting times.

Integrating (\ref{2}), one gets
\begin{equation}\label{ts}
t(s)=\int_0^sds'\mu(s-s')\tau(s')
\end{equation}
with $\mu(s-s')=\int_{s'}^s ds''M(s''-s')$. Then the Laplace transform $(t\rightarrow \lambda)$ of the PDF $p(t,s)$ of $t(s)$ is \cite{ChechkinHofmannSokolov:2009}
\begin{equation}
p(\lambda,s)=\int_0^\infty dt e^{-\lambda t}p(t,s)=e^{-\lambda^\alpha \phi(s)},
\end{equation}
where $\phi(s)=\int_0^s\mu^\alpha(s')ds'$. Define the inverse of the stochastic process $t(s)$ as
\begin{equation}
s(t)=t^{-1}(s)=\inf\{s>0: t(s)>t\}.
\end{equation}
Using the monotonicity of $t(s)$, i.e., $s_2>s_1\Rightarrow t(s_2)>t(s_1)$, one arrives at \cite{BauleFriedrich:2005}
\begin{equation}\label{Theta}
\Theta(s-s(t))=1-\Theta(t-t(s)),
\end{equation}
where $\Theta(x)$ is the Heaviside function, $\Theta(x)=1$ for $x>0$, $\Theta(x)=0$ for $x<0$ and $\Theta(x=0)=1/2$. The PDF $h(s,t)$ of $s(t)$ is therefore can be represented as
\begin{equation}
\begin{split}
h(s,t)=\langle \delta(s-s(t))\rangle&=\frac{\partial}{\partial s}\langle \Theta(s-s(t))\rangle\\
&=-\frac{\partial}{\partial s}\langle \Theta(t-t(s))\rangle.
\end{split}
\end{equation}
Then the Laplace transform of the PDF $h(s,t)$ is
\begin{equation}\label{onepoint}
h(s,\lambda)=-\frac{\partial}{\partial s}\frac{1}{\lambda}p(\lambda,s)=\lambda^{\alpha-1}\phi'(s)p(\lambda,s).
\end{equation}
Detailed derivations of above formulas can be found in \cite{ChechkinHofmannSokolov:2009}.

For the rest of this section, we deduce the two-point joint PDFs of the stochastic process $t(s)$ and its inverse process $s(t)$ in Laplace space $(t_1\rightarrow\lambda_1, t_2\rightarrow\lambda_2)$, denoted as $p(\lambda_2,s_2;\lambda_1,s_1)$ and $h(s_2,\lambda_2;s_1,\lambda_1)$, respectively. From (\ref{Theta}), we get
\begin{equation}
\begin{split}
&\langle\Theta(s_2-s(t_2))\Theta(s_1-s(t_1))\rangle\\
&~~=\langle[1-\Theta(t_2-t(s_2))][1-\Theta(t_1-t(s_1))]\rangle\\
&~~=1-\langle \Theta(t_2-t(s_2))\rangle-\langle \Theta(t_1-t(s_1))\rangle\\
&~~~~~+\langle\Theta(t_2-t(s_2))\Theta(t_1-t(s_1)) \rangle.
\end{split}
\end{equation}
Then the two-point joint PDF of the inverse process $s(t)$ is
\begin{equation}\label{ADD1}
\begin{split}
h&(s_2,t_2;s_1,t_1) \\
&=\langle \delta(s_2-s(t_2))\delta(s_1-s(t_1))\rangle\\
&=\frac{\partial}{\partial s_1}\frac{\partial}{\partial s_2}\langle \Theta(t_2-t(s_2))\Theta(t_1-t(s_1))\rangle\\
&=\frac{\partial}{\partial s_1}\frac{\partial}{\partial s_2} \int_0^{t_2}dt'_2\int_0^{t_1}dt'_1\langle\delta(t'_2-t(s_2))\delta(t'_1-t(s_1))\rangle.
\end{split}
\end{equation}
Utilizing the Laplace transform method for \eqref{ADD1}, finally, the relationship between the two-point joint PDF $h(s_2, t_2; s_1, t_1)$ of $s(t)$ and $p(t_2,s_2;t_1,s_1)$ of $t(s)$ in Laplace space becomes \cite{BauleFriedrich:2005}
\begin{equation}\label{twopoint}
\begin{split}
h(s_2,\lambda_2;s_1,\lambda_1)&=\frac{\partial}{\partial s_1}\frac{\partial}{\partial s_2}\frac{1}{\lambda_1\lambda_2}\Bigg\langle e^{-\lambda_2t(s_2)}e^{-\lambda_1t(s_1)}\Bigg\rangle \\
&=\frac{\partial}{\partial s_1}\frac{\partial}{\partial s_2} \frac{1}{\lambda_1\lambda_2}p(\lambda_2,s_2;\lambda_1,s_1).
\end{split}
\end{equation}

According to the above discussions, to obtain the two-point joint PDF of the inverse process $s(t)$, one should first get the two-point joint PDF of the stochastic process $t(s)$. Using the expression of $t(s)$ in (\ref{ts}), the two-point joint PDF of the stochastic process $t(s)$ in Laplace space is
\begin{equation}\label{p}
\begin{split}
&p(\lambda_2,s_2;\lambda_1,s_1)\\
&~~=\mathcal{L}_{t_1\rightarrow\lambda_1, t_2\rightarrow\lambda_2}[\langle \delta(t_2-t(s_2))\delta(t_1-t(s_1))\rangle]\\
&~~=\left\langle \exp\left\{-\lambda_2\int_0^{s_2}ds'_2\tau(s'_2)\mu(s_2-s'_2)\right\}\right. \\
&~~~~~~\left.\cdot\exp\left\{-\lambda_1\int_0^{s_1}ds'_1\tau(s'_1)\mu(s_1-s'_1)\right\}\right\rangle.
\end{split}
\end{equation}
To calculate the ensemble average in \eqref{p}, we need to split one of the integrals into two parts and make the best use of the properties of the $\alpha$-stable totally skewed L\'{e}vy process $t_0(s)=\int_0^s ds'\tau(s')$ which has stationary and independence increments, together with the Laplace transform of its PDF $L_\alpha(t_0,s)$, i.e.,
\begin{equation}\label{character}
\mathcal{L}_{t_0\rightarrow \lambda_0}[L_\alpha(t_0,s)]=e^{-\lambda_0^\alpha s},~~~~0<\alpha<1.
\end{equation}
More precisely, assuming $s_1<s_2$, (\ref{p}) can be rewritten as
\begin{equation}\label{ADD3}
\begin{split}
&p(\lambda_2,s_2;\lambda_1,s_1)\\
&~~=\left\langle \exp\left\{-\lambda_2\int_0^{s_1}ds'_2\tau(s'_2)\mu(s_2-s'_2)\right\}\right. \\
&~~~~~~\left.\cdot\exp\left\{-\lambda_1\int_0^{s_1}ds'_1\tau(s'_1)\mu(s_1-s'_1)\right\}\right\rangle \\
&~~~~~~\cdot\left\langle \exp\left\{-\lambda_2\int_{s_1}^{s_2}ds'\tau(s')\mu(s_2-s')\right\}\right\rangle \\
&~~=\exp\left\{-\int_0^{s_1}ds'[\lambda_1\mu(s_1-s')+\lambda_2\mu(s_2-s')]^\alpha\right\} \\
&~~~~~~\cdot\exp\left\{-\int_0^{s_2-s_1}ds'[\lambda_2\mu(s_2-s_1-s')]^\alpha\right\},
\end{split}
\end{equation}
where the detailed derivations of the last step are shifted to the Appendix.
Finally, we obtain the two-point joint PDF of $t(s)$ in Laplace space
\begin{equation}
\begin{split}
p&(\lambda_2,s_2;\lambda_1,s_1)\\
&=\Theta(s_2-s_1)\exp\left\{-\int_0^{s_2-s_1}ds'[\lambda_2\mu(s_2-s_1-s')]^\alpha\right\} \\
&~~~~\cdot\exp\left\{-\int_0^{s_1}ds'[\lambda_1\mu(s_1-s')+\lambda_2\mu(s_2-s')]^\alpha\right\}\\
&~~+\Theta(s_1-s_2)\exp\left\{-\int_0^{s_1-s_2}ds'[\lambda_1\mu(s_1-s_2-s')]^\alpha\right\}\\
&~~~~\cdot\exp\left\{-\int_0^{s_2}ds'[\lambda_1\mu(s_1-s')+\lambda_2\mu(s_2-s')]^\alpha\right\} .
\end{split}
\end{equation}
Taking $\mu(s)=1$ (i.e., $M(s)=\delta(s)$ in \eqref{2}), the two-point joint PDF of stochastic process $t(s)$ reduces to
\begin{equation*}
\begin{split}
p(\lambda_2,s_2;\lambda_1,s_1)&=\Theta(s_2-s_1)e^{-s_1(\lambda_1+\lambda_2)^\alpha}e^{-(s_2-s_1)\lambda_2^\alpha}\\
&~~+\Theta(s_1-s_2)e^{-s_2(\lambda_1+\lambda_2)^\alpha}e^{-(s_1-s_2)\lambda_1^\alpha},
\end{split}
\end{equation*}
which is identical to the uncorrelated case in \cite{BauleFriedrich:2005}.
By formula (\ref{twopoint}), after some calculations, one arrives at the two-point joint PDF $h(s_2,t_2;s_1,t_1)$ of inverse process $s(t)$ in Laplace space
\begin{equation}\label{h}
\begin{split}
h&(s_2,\lambda_2;s_1,\lambda_1)\\
&=\Theta(s_2-s_1)\frac{1}{\lambda_1\lambda_2}\left(\frac{\partial f_1}{\partial s_1}\frac{\partial f_1}{\partial s_2}-\frac{\partial ^2 f_1}{\partial s_1\partial s_2}\right)e^{-f_1}\\
&~~+\Theta(s_1-s_2)\frac{1}{\lambda_1\lambda_2}\left(\frac{\partial f_2}{\partial s_1}\frac{\partial f_2}{\partial s_2}-\frac{\partial ^2 f_2}{\partial s_1\partial s_2}\right)e^{-f_2},
\end{split}
\end{equation}
where
\begin{equation*}
\begin{split}
f_1&=\int_0^{s_1}ds'[\lambda_1\mu(s_1-s')+\lambda_2\mu(s_2-s')]^\alpha\\
&~~~+\int_0^{s_2-s_1}ds'
[\lambda_2\mu(s_2-s_1-s')]^\alpha,\\
f_2&=\int_0^{s_2}ds'[\lambda_1\mu(s_1-s')+\lambda_2\mu(s_2-s')]^\alpha\\
&~~~+\int_0^{s_1-s_2}ds'
[\lambda_1\mu(s_1-s_2-s')]^\alpha.
\end{split}
\end{equation*}

\section{L\'{e}vy-walk-type model with correlated waiting times}\label{three}

Since the Langevin picture of the CTRW with heavy-tailed distributed waiting times was introduced by Fogedby \cite{Fogedby:1994}, it seems that the subordinator (especially $\alpha$-stable totally skewed L\'{e}vy process) is more often  applied to describe subdiffusion processes. In fact, the subordinator can also be used to characterize superdiffusion processes using the underdamped Langevin equation, which corresponds to a random walk with random velocity proposed by \cite{EuleZaburdaevFriedrichGeisel:2012}. In \cite{EuleZaburdaevFriedrichGeisel:2012}, the stochastic process described by the coupled Langevin equation exhibits ballistic diffusion when the $\alpha$-stable subordinator with $0<\alpha<1$ is applied. Now, we extend the waiting times to be correlated and modify the subordinator to be the correlated time process $t(s)$ in (\ref{ts}). Finally, the Langevin picture of the L\'{e}vy-walk-type model with correlated waiting times can be written as:
\begin{equation}\label{model}
\begin{split}
\frac{dx(t)}{dt}&=v(t),\\
\frac{dv(s)}{ds}&=-\gamma v(s)+\eta(s),\\
\frac{dt(s)}{ds}&=\int_0^sds'M(s-s')\tau(s').
\end{split}
\end{equation}
Here the memory function $M(s)$ is taken to be power-law function $M(s)=s^{\beta-1}/\Gamma(\beta)$ with $0<\beta<1$. With the increase of $\beta$, the impact of historical waiting time on the current waiting time becomes larger. That is to say, the correlation of waiting times becomes stronger. In \eqref{model}, $v(s)$ is the stochastic process of velocity with operation time $s$, $\tau(s)$ and $\eta(s)$ are the $\alpha$-stable L\'{e}vy noise and the white Gaussian noise respectively, which have been mentioned in the last section, and $\gamma$ is the friction coefficient. The process of velocity with respect to physical time is defined as $v(t):=v(s(t))$, where $s(t)$ is the inverse process of $t(s)$. Then the stochastic process of position with respect to physical time that we are interested in is
\begin{equation}
x(t)=\int_0^tdt'v(s(t')).
\end{equation}

As is well known, sometimes a stochastic process can be characterized by both Langevin equations and CTRWs, which are two equivalent models after taking the scaling limit. Then the MSD of a stochastic process can be obtained from both of the two frameworks. But for the process with correlated waiting times, the master equation in CTRWs becomes difficult to obtain; hence it is not easy to get the MSD in this way, while the MSD and the correlation function can be obtained from the solutions of the Langevin equations directly. This is a relative advantage of the Langevin equations.

One-point and two-point PDFs of the stochastic process $v(t)$ is denoted as $f(v,t)$ and $f(v_2,t_2;v_1,t_1)$, respectively. Using the independence of the processes $v(s)$ and $s(t)$ leads to
\begin{equation}
\begin{split}
  f(v,t)&=\langle \delta(v-v(t))\rangle \\
  &=\int_0^\infty ds\langle \delta(v-v(s))\rangle\langle \delta(s-s(t))\rangle.
\end{split}
\end{equation}
Then $f(v,t)$ can be written as
\begin{equation}\label{f_v}
f(v,t)=\int_0^\infty dsf_0(v,s)h(s,t),
\end{equation}
where $f_0(v,s)$ is the one-point PDF of stochastic process $v(s)$.
Similarly, the two-point joint PDF $f(v_2,t_2;v_1,t_1)$ of $v(t)$ can be obtained through two-point joint PDF $f_0(v_2,s_2;v_1,s_1)$ of stochastic process $v(s)$,
\begin{equation}\label{f_v12}
\begin{split}
&f(v_2,t_2;v_1,t_1)\\
&~~=\int_0^\infty ds_1\int_0^\infty ds_2f_0(v_2,s_2;v_1,s_1)h(s_2,t_2;s_1,t_1).
\end{split}
\end{equation}
Multiplying $v$ on both sides of (\ref{f_v}) and integrating with respect to $v$, then the first moment of stochastic process $v(t)$ in Laplace space ($t\rightarrow \lambda$) is
\begin{equation}
\langle v(\lambda)\rangle=\int_0^\infty ds\langle v(s)\rangle h(s,\lambda).
\end{equation}
Similarly, multiplying $v_1v_2$ on both sides of  (\ref{f_v12}), doing integration with respect to $v_1$ and $v_2$, respectively, and making Laplace transform $(t_1\rightarrow\lambda_1, t_2\rightarrow\lambda_2)$, one arrives at the correlation function of $v(t)$ in Laplace space
\begin{equation}\label{vv}
\begin{split}
&\langle v(\lambda_1)v(\lambda_2)\rangle\\
&~~=\int_0^\infty ds_1\int_0^\infty ds_2\langle v(s_1)v(s_2)\rangle h(s_2,\lambda_2;s_1,\lambda_1).
\end{split}
\end{equation}

We will consider the stochastic process with strong correlated waiting times, i.e., $\beta=1$ in the memory function $M(s)=s^{\beta-1}/\Gamma(\beta)$, in subsections Sec. \ref{IIIA} and Sec. \ref{IIIB}; and in Sec. \ref{IIIC} we focus on the more general case $0<\beta<1$.


\subsection{Strong correlation of waiting times with $\beta=1$ and friction coefficient $\gamma\neq0$}\label{IIIA}
In the case of $\beta=1$, i.e., $\mu(s)=s$, inserting it into formula (\ref{onepoint}) and (\ref{h}), one arrives at the one-point PDF $h(s,t)$ of the inverse process $s(t)$ in Laplace space
\begin{equation}\label{hs}
h(s, \lambda)=\lambda^{\alpha-1}s^\alpha e^{-\frac{\lambda^\alpha}{1+\alpha}s^{1+\alpha}},
\end{equation}
and the two-point joint PDF $h(s_2, t_2; s_1, t_1)$ of $s(t)$ in Laplace space
\begin{equation}\label{h_s12}
\begin{split}
h&(s_2, \lambda_2; s_1,\lambda_1)\\
&=\Theta(s_2-s_1) \Bigg( \frac{(\Phi_1^\alpha-\Phi_2^\alpha)^2}
{(\lambda_1+\lambda_2)^2}
-\frac{\Phi_1^\alpha(\Phi_1^\alpha-\Phi_2^\alpha)}{\lambda_2(\lambda_1+\lambda_2)}\\
&~~+\frac{\alpha(\Phi_1^{\alpha-1}-\Phi_2^{\alpha-1})}{\lambda_1+\lambda_2}\Bigg)
\exp\left[-\frac{\lambda_2\Phi_2^{\alpha+1}+\lambda_1\Phi_1^{\alpha+1}}
{\lambda_2(\lambda_1+\lambda_2)(\alpha+1)}\right]\\
&~~+\Theta(s_1-s_2)  \Bigg(\frac{(\Phi_3^\alpha-\Phi_2^\alpha)^2}
{(\lambda_1+\lambda_2)^2}
-\frac{\Phi_3^\alpha(\Phi_3^\alpha-\Phi_2^\alpha)}{\lambda_1(\lambda_1+\lambda_2)}\\
&~~+\frac{\alpha[\Phi_3^{\alpha-1}-\Phi_2^{\alpha-1}]}{\lambda_1+\lambda_2} \Bigg)
\exp\left[-\frac{\lambda_1\Phi_2^{\alpha+1}+\lambda_2\Phi_3^{\alpha+1}}
{\lambda_1(\lambda_1+\lambda_2)(\alpha+1)}\right],
\end{split}
\end{equation}
where $\Phi_1=\lambda_2(s_2-s_1)$, $\Phi_2=\lambda_1s_1+\lambda_2s_2$, and $\Phi_3=\lambda_1(s_1-s_2)$.

\begin{figure}
\centering
\includegraphics[scale=0.5]{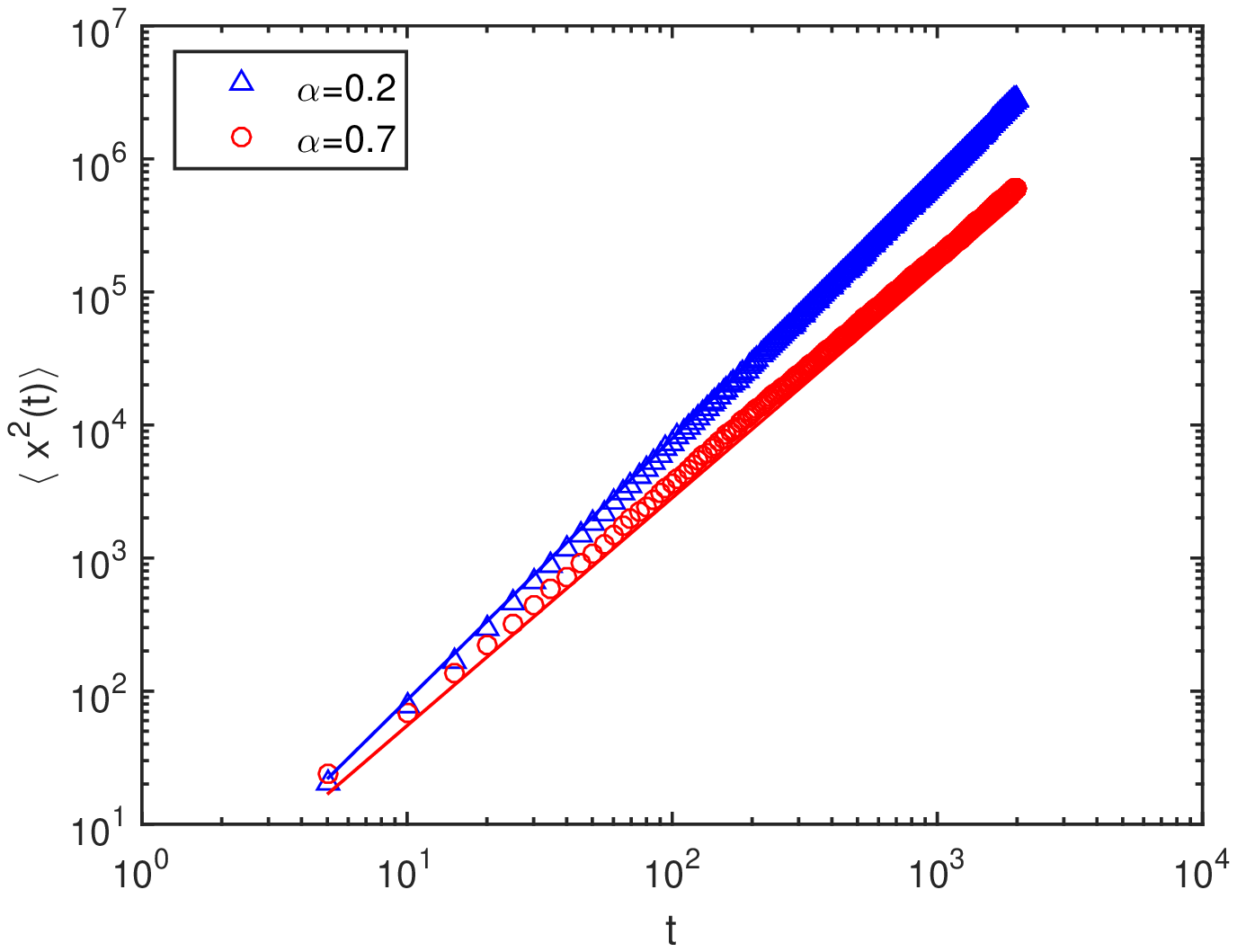}
\caption{MSD of the stochastic process described by the Langevin picture of L\'{e}vy-walk-type model with strong correlated waiting times in the case of friction coefficient $\gamma\neq0$.  Solid lines represent the theoretical values (\ref{MSD}) and the markers are simulation results. The  parameters are respectively taken as $\gamma=1, D_v=1$, $\beta=1$, and the initial velocity is chosen to be $v_0=0$. }\label{MSD_gama_1}
\end{figure}

Considering the effect of friction, i.e., $\gamma\neq0$, using Laplace transform technique, the solution of the second equation in (\ref{model}) is
\begin{equation}\label{vs}
v(s)=\int_0^sds'\eta(s')e^{-\gamma (s-s')}+v_0e^{-\gamma s}.
\end{equation}
Taking initial velocity $v_0=0$, the first moment and correlation function of velocity process $v(s)$ are $\langle v(s)\rangle=0$ and
\begin{equation}\label{vsvs}
\langle v(s_1)v(s_2)\rangle=\frac{D_v}{\gamma}\bigg(e^{-\gamma|s_1-s_2|}-e^{-\gamma(s_1+s_2)}\bigg).
\end{equation}
Therefore, the first moment of velocity process $v(t)$ with respect to physical time $t$ is zero. As for the second moment, using
\begin{equation}
\langle v^2(\lambda)\rangle=\int_0^\infty ds\langle v^2(s)\rangle h(s,\lambda),
\end{equation}
and (\ref{hs}), after taking the inverse Laplace transform,
one can easily obtain that $\langle v^2(t)\rangle\simeq D_v/\gamma$ for long times. However, it cannot be used to obtain the MSD of process $x(t)$.
Substituting the correlation function of $v(s)$ and (\ref{h_s12}) into (\ref{vv}), after some complex calculations, the correlation function of $v(t)$ in Laplace space for small $\lambda_1$ and $\lambda_2$ is obtained as
\begin{equation}\label{v_lambda}
\langle v(\lambda_1)v(\lambda_2)\rangle\simeq K_1\frac{\lambda_1^{\alpha-1}+\lambda_2^{\alpha-1}}{(\lambda_1+\lambda_2)^{1+\tilde\alpha}},
\end{equation}
where $\tilde\alpha=\frac{\alpha}{1+\alpha}$ and $K_1=\frac{D_v\Gamma(1+\alpha)\Gamma(\frac{1}{1+\alpha})}
{\gamma^{1+\alpha}(1+\alpha)^{\tilde\alpha}2^{\tilde\alpha}}$.
Taking the inverse Laplace transform on (\ref{v_lambda}) leads to the correlation function of $v(t)$ for long times
\begin{equation}\label{corV}
\langle v(t_1)v(t_2)\rangle\simeq K_2(t_2-t_1)^{-\alpha}t_1^{\tilde\alpha},
\end{equation}
where $K_2=K_1\frac{1}{\Gamma(1-\alpha)}\frac{1}{\Gamma(1+\tilde\alpha)}$ and $t_2>t_1$.
Known the correlation function of velocity process, the correlation function of the stochastic process $x(t)$ that we are interested in has the expression
\begin{equation}\label{cov}
\begin{split}
\langle &x(t_1)x(t_2)\rangle \\
&=\int_0^{t_1}dt'_1\int_0^{t_2}dt'_2\langle v(t'_1)v(t'_2)\rangle \\
&\simeq \Theta(t_2-t_1)K_2\Bigg[\frac{1}{2-\alpha\tilde\alpha}B(1+\tilde\alpha,1-\alpha)
t_1^{2-\alpha\tilde\alpha}\\
&~~~+\frac{1}{1-\alpha}B(1+\tilde\alpha,2-\alpha;t_1/t_2)
t_2^{2-\alpha\tilde\alpha}\Bigg]\\
&~~~+\Theta(t_1-t_2)K_2\Bigg[\frac{1}{2-\alpha\tilde\alpha}B(1+\tilde\alpha,1-\alpha)
t_2^{2-\alpha\tilde\alpha}\\
&~~~+\frac{1}{1-\alpha}B(1+\tilde\alpha,2-\alpha;t_2/t_1)
t_1^{2-\alpha\tilde\alpha}\Bigg],
\end{split}
\end{equation}
where $B(a,b;z)=\int_0^z d\tau\tau^{a-1}(1-\tau)^{b-1}$ is the incomplete Beta function \cite{AbramowitzStegun:1972}. It can be seen that the correlation function $\langle x(t_1)x(t_2)\rangle$ cannot be expressed as a function of time difference $|t_1-t_2|$, meaning that the process described by model (\ref{model}) is nonstationary. When $t_1=t_2$, the MSD of stochastic process $x(t)$ for long time $t$ is
\begin{equation}\label{MSD}
\langle x^2(t)\rangle\simeq \frac{2K_2}{1-\alpha}B\left(1+\tilde\alpha,2-\alpha\right)t^{2-\alpha\tilde\alpha},
\end{equation}
which shows that the L\'{e}vy-walk-type model with strong correlated waiting times (\ref{model}) ($\beta=1$) exhibits sub-ballistic superdiffusion phenomenon. The simulation results together with analytical results (\ref{MSD}) for different $\alpha$ are presented in Fig. \ref{MSD_gama_1} and it can be seen that the simulation results are consistent with the analytical ones. With the increase of $\alpha$, the diffusion rate slows down. Comparing with the Langevin description of L\'{e}vy-walk-type model with i.i.d. waiting times in \cite{EuleZaburdaevFriedrichGeisel:2012}, which exhibits ballistic diffusion phenomenon for long times, we find that the correlated waiting times slows down the diffusion of particles.

In the effect of friction ($\gamma\neq0$), for long times, the variance of velocity $v(t)$ of the stochastic process described by model (\ref{model}) becomes a constant, i.e., $\langle v^2(t)\rangle\simeq D_v/\gamma$, which implies the velocity process could reach a steady state. In this case, the model (\ref{model}) can be regarded as the Langevin picture of the L\'{e}vy walk model where a particle moves ballistically for a random but correlated time and then changes direction but keeps the same magnitude of velocity.
Intuitively, the correlation makes waiting time grow longer. 
 That is to say, the time that particles spend at the same velocity and same direction becomes longer, and thus lengthens the displacement of the particle. It seems that the diffusion of the particles is enhanced compared with the stochastic process described by L\'{e}vy walk model with uncorrelated waiting times.  Nevertheless, in (\ref{model}), the correlation of waiting times lengthens the times of all the steps except the one of the initial step, and the time in the  current step is longer than the time in the immediately previous step.
 When a particle finishes its current step, it will make new choice for its moving direction. The longer waiting time in the next step just makes possibly the opposite distance become longer, which actually suppresses the diffusion. Therefore, the stochastic process described by Langevin system (\ref{model}) with $\beta=1$ displays sub-ballistic diffusion phenomenon as Fig. \ref{MSD_gama_1} shows, which is slower than the L\'{e}vy walk model with independent power-law distributed waiting times.
Besides, it can be noted that for fixed $\beta=1$, with the increase of $\alpha$, the diffusion rate of the particles in our model slows down. Since the larger $\alpha$ reduces the probability of longer waiting times, which leads to the shorter moving distance for the particles with constant velocity. It is different from the L\'{e}vy walk model with uncorrelated waiting times, which presents ballistic diffusion phenomenon independent of the value of $\alpha$.

Using (\ref{cov}) and (\ref{MSD}), as well as the asymptotic expression of the incomplete Beta function for small $z$, i.e., $B(a,b;z)\simeq z^a/a$,
for fixed $t_1$ and $t_2\rightarrow \infty$, the correlation coefficient $r[x(t_1),x(t_2)]$ of the stochastic process $x(t)$ could be obtained as
\begin{equation}\label{co1}
r[x(t_1),x(t_2)]=\frac{\langle x(t_1)x(t_2)\rangle}{\sqrt{\langle x^2(t_1)\rangle\langle x^2(t_2)\rangle}}
\simeq K_3 \left(\frac{t_1}{t_2}\right)^{\tilde\alpha(2+\alpha)/2},
\end{equation}
where $K_3=\frac{2(1+\alpha)-\alpha^2}{2(1+\alpha-2\alpha^2)B(1+\tilde\alpha,1-\alpha)}$. That is to say, the process $x(t)$ described by model (\ref{model}) ($\beta=1$) is long-range dependent \cite{WylomanskaKumarPoloczanskiVellaisamy:2016} since $0<\tilde\alpha(2+\alpha)/2<1$. For L\'{e}vy walk model with uncorrelated waiting time obeying power-law distribution, which displays ballistic diffusion phenomenon in the case of $0<\alpha<1$, we take use of the correlation function and MSD of $x(t)$ in \cite{FroembergBarkai:2013}, and obtain the correlation coefficient 
\begin{equation}\label{LW}
r_0[x(t_1),x(t_2)]=\frac{\langle x(t_1)x(t_2)\rangle}{\sqrt{\langle x^2(t_1)\rangle\langle x^2(t_2)\rangle}}
\simeq K_4 \left(\frac{t_1}{t_2}\right)^\alpha,
\end{equation}
where $K_4=\frac{\sin\pi \alpha}{\pi\alpha(1-\alpha)(1-\alpha^2)}$. Comparing (\ref{co1}) with (\ref{LW}), the stronger correlation of the stochastic process $x(t)$ is found in the L\'{e}vy-walk-type model with correlated waiting times in model (\ref{model}) since $\tilde\alpha(2+\alpha)/2<\alpha$.

Next, we study the ergodic property of the stochastic process described by the Langevin system (\ref{model}) in the case of $\beta=1$, i.e., L\'{e}vy-walk-type model with strong correlated waiting times. The ergodicity means that the ensemble average and time average for an observable are identical in the limit of long measure times.  The time averaged MSD is defined as
\begin{equation}\label{timeaverage}
\overline{\delta^2}=\frac{1}{T-\Delta}\int_0^{T-\Delta}dt[x(t+\Delta)-x(t)]^2,
\end{equation}
where $\Delta$ is the lag time, and $T$ is measurement time. We emphasize that the lag time $\Delta$ separating the displacement between trajectory points is much shorter than measurement time $T$. Weak ergodicity breaking says that the particles could explore the whole phase space but with a diverging time scale exceeding the measurement time \cite{KlagesRadonsSokolov:2008,FroembergBarkai:2013}. This phenomenon can be observed in many systems, typically with power-law distributed waiting times, such as glassy dynamics \cite{Bouchaud:1992}, diffusion of molecules in cell environment \cite{WeigelSimonTamkunKrapf:2011}. In \cite{FroembergBarkai:2013}, weak ergodicity breaking of the L\'{e}vy walk model is detected, where the particle's velocity alternates at the power-law distributed random times with power-law exponent $0<\alpha<1$; for a constant velocity $v$, the mean of time averaged MSD of the stochastic process is $\langle\overline{\delta^2}\rangle\simeq v^2\Delta^2$ for $\Delta/T\ll 1$, which is different from the ensemble averaged MSD $\langle x^2\rangle \simeq v^2(1-\alpha)\Delta^2$ by a constant prefactor; this phenomenon is named ultraweak ergodicity breaking in \cite{GodecMetzler:2013}.

\begin{figure}
\begin{minipage}{0.35\linewidth}
  \centerline{\includegraphics[scale=0.58]{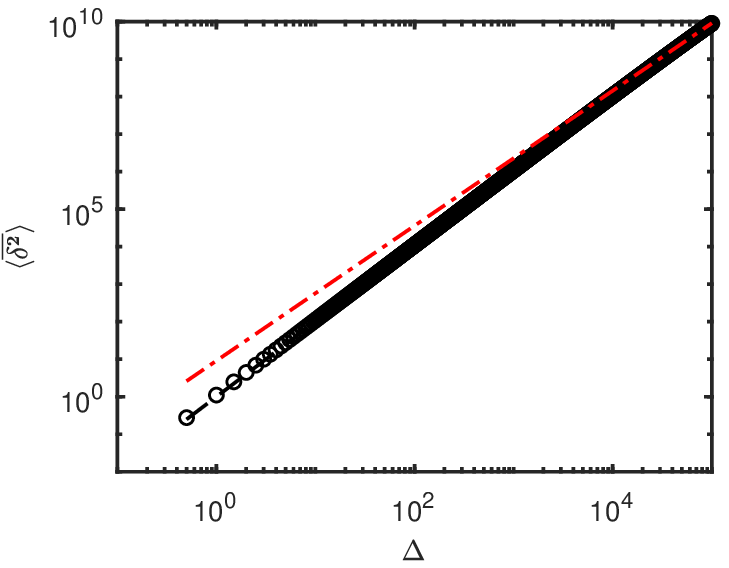}}
  \centerline{(a)}
\end{minipage}
\hspace{1cm}
\begin{minipage}{0.35\linewidth}
  \centerline{\includegraphics[scale=0.58]{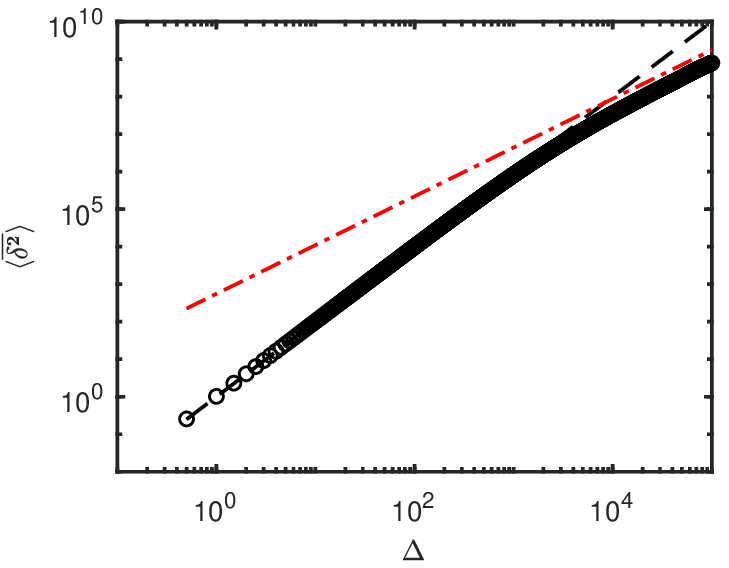}}
  \centerline{(b)}
\end{minipage}
\caption{Mean value of time averaged MSD $\langle \overline{\delta^2}\rangle$ of the stochastic process described by the L\'{e}vy-walk-type model with strong correlated waiting times for the measurement time $T=2\times 10^5$, $\gamma=1$, $D_v=1$, $\beta=1$, $\alpha=0.2$ (left panel) or $\alpha=0.7$ (right panel), and the initial velocity $v_0=0$. The red dashed dot lines and the black dashed lines show the theoretical results of $\langle \overline{\delta^2}\rangle$ for large $\Delta$ (\ref{meantimeaverage}) and small $\Delta$ ($\langle \overline{\delta^2}\rangle\simeq D_v/\gamma\Delta^2$), respectively, which coincide with the simulation results represented by the circle-markers. }\label{ergodicity}
\end{figure}

To investigate the ergodic property of the stochastic process described by model (\ref{model}), we insert the correlation function (\ref{cov}) together with the MSD (\ref{MSD}) of the process $x(t)$ into $\langle [x(t+\Delta)-x(t)]^2\rangle$, and then obtain
\begin{equation}\label{ADD2}
\begin{split}
&\langle[x(t+\Delta)-x(t)]^2\rangle\\
&~~=\langle x^2(t+\Delta)\rangle+\langle x^2(t)\rangle-2\langle x(t+\Delta)x(t)\rangle\\
&~~\simeq \frac{2}{(1-\alpha)(2-\alpha)}\Delta^{2-\alpha}(t+\Delta)^{\tilde\alpha}
\end{split}
\end{equation}
for large $t$. This result not only depends on the lag time $\Delta$ but also the time $t$, reflecting the aging phenomenon in the system by regarding $t$ as the aging time $t_a$.
Substituting \eqref{ADD2} into the mean of \eqref{timeaverage}, we obtain the mean value of time averaged MSD $\langle \overline{\delta^2}\rangle$ in the asymptotic $\Delta\ll T$,
\begin{equation}\label{meantimeaverage}
\begin{split}
\langle\overline{\delta^2}\rangle&=\frac{1}{T-\Delta}\int_0^{T-\Delta}dt\langle[x(t+\Delta)-x(t)]^2\rangle \\[3pt]
&\simeq K_5\Delta^{2-\alpha}T^{\tilde\alpha},
\end{split}
\end{equation}
with $K_5=\frac{2}{(1-\alpha)(2-\alpha)(1+\tilde\alpha)}$.
Comparing \eqref{MSD} with \eqref{meantimeaverage}, it can be noted that the ensemble averaged MSD and the time averaged MSD are different, implying the weak ergodicity breaking of the process $x(t)$. The simulation results together with theoretical ones of the mean of time averaged MSD for different $\alpha$ are shown in Fig. \ref{ergodicity}. For small $\Delta$, the particle keeps the same velocity and does not change its moving direction, so $\langle[x(t+\Delta)-x(t)]^2\rangle= \langle v^2\rangle\Delta^2$ with $\langle v^2\rangle\simeq D_v/\gamma$ for long times, and inserting it into (\ref{timeaverage}) yields $\langle\overline{\delta^2}\rangle\simeq D_v/\gamma\Delta^2$. While with the increase of $\Delta$, the mean value of time averaged MSD becomes $\langle \overline{\delta^2}\rangle\simeq K_5\Delta^{2-\alpha}T^{\tilde\alpha}$ in (\ref{meantimeaverage}). The theoretical results perfectly coincide with the simulation ones.

We have known that the stochastic process described by L\'{e}vy walk model with independent {heavy-tailed} waiting times ($0<\alpha<1$) is ultraweak ergodicity breaking \cite{FroembergBarkai:2013,GodecMetzler:2013} with the ergodicity breaking parameter
\begin{equation}\label{EB1}
\begin{split}
\mathcal{EB}=\frac{\langle\overline{\delta^2}(\Delta)\rangle}{\langle x^2(\Delta)\rangle}\simeq\frac{1}{1-\alpha}.
\end{split}
\end{equation}
For the L\'{e}vy-walk-type model (\ref{model}), which exhibits weak ergodicity breaking due to the strong correlation of waiting times, has ergodicity breaking parameter
\begin{equation}\label{EB2}
\begin{split}
\mathcal{EB}=\frac{\langle\overline{\delta^2}(\Delta)\rangle}{\langle x^2(\Delta)\rangle}\propto\left(\frac{T}{\Delta}\right)^{\tilde{\alpha}}
\end{split}
\end{equation}
with $\Delta/T\ll1$. Comparing the two ergodicity breaking parameters, (\ref{EB1}) and (\ref{EB2}), one can find a more strong ergodicity breaking phenomenon in our model (\ref{model}) with strong correlated waiting times, since the exponent in ergodicity breaking parameter $\tilde{\alpha}>0$, which means a more obvious deviation between time averaged MSD and ensemble averaged MSD appears in (\ref{EB2}) comparing with (\ref{EB1}).

At the end of this subsection, let us show the self-similar property of the processes $t(s)$, $s(t)$, $v(t)$, and $x(t)$.
Integrating the expression of $t(s)$ in (\ref{model}) and letting $\beta=1$, one has
\begin{equation}
t(s)=\int_0^sds'\int_0^{s'}ds''\tau(s'').
\end{equation}
Recall that the $\alpha$-stable L\'{e}vy process is $1/\alpha$-self-similar \cite{Applebaum:2009}. Then for any $a>0$,
\begin{equation}
\begin{split}
&t(as)=\int_0^{as}ds'\int_0^{s'}ds''\tau(s'')\\
&~~~~~~\overset{d}{=}a^{1+\frac{1}{\alpha}}\int_0^sds'\int_0^{s'}ds''\tau(s'')\\
&~~~~~~=a^{1+\frac{1}{\alpha}}t(s),
\end{split}
\end{equation}
where $\overset{d}{=}$ denotes identical distribution.
Therefore, the stochastic process $t(s)$ is $1/\tilde\alpha$-self-similar. Applying the method in \cite{MagdziarzMetzlerSzczotkaZebrowski:2012_2} leads to
\begin{equation}
\begin{split}
&P[t^{-1}(as)\leq\tau]\\
&~~=P[as\leq t(\tau)]=P[s\leq a^{-1}t(\tau)]=P[s\leq t(a^{-\tilde\alpha}\tau)]\\
&~~=P[t^{-1}(s)\leq a^{-\tilde\alpha}\tau]=P[a^{\tilde\alpha}t^{-1}(s)\leq\tau],
\end{split}
\end{equation}
which means the inverse process $s(t)$ is $\tilde\alpha$-self-similar.

For process $v(t)$ with $v_0=0$, we first rewrite $v(s)$ in (\ref{vs}) as $v_\gamma(s)$, and then obtain
\begin{equation}
\begin{split}
&v_{\gamma}(as)=\int_0^{as}B(ds')e^{-\gamma (as-s')}\\
&~~~~~~~~\overset{d}{=}a^{\frac{1}{2}}\int_0^sB(ds')e^{-\gamma a(s-s')}\\
&~~~~~~~~=a^{\frac{1}{2}}v_{\gamma a}(s).
\end{split}
\end{equation}
The scaling property of the process $v_\gamma(t)$ can be obtained through the self-similarity of $s(t)$:
\begin{equation}
\begin{split}
v_\gamma(at)=v_\gamma(s(at))\overset{d}{=}v_\gamma(a^{\tilde\alpha} s(t))\overset{d}{=}a^{\tilde\alpha/2}v_{\gamma a^{\tilde\alpha}}(t).
\end{split}
\end{equation}
Similarly, the scaling property of the stochastic process $x_\gamma(t)$, which is the integration of velocity process $v_\gamma(t)$, is
\begin{equation}
x_\gamma(at)\overset{d}{=}a^{1+\tilde\alpha/2}x_{\gamma a^{\tilde\alpha}}(t).
\end{equation}
In the case of neglecting friction, i.e., $\gamma=0$, the stochastic process $x(t)$ has self-similar property
\begin{equation}\label{self}
x(at)\overset{d}{=}a^{1+\tilde\alpha/2}x(t),
\end{equation}
which could be a convenient way to obtain the $n$-th moment of $x(t)$, seeing (\ref{moments}) in the next subsection for the non-friction case.

\subsection{Strong correlation of waiting times with $\beta=1$ and friction coefficient $\gamma=0$}\label{IIIB}

\begin{figure}
\centering
\includegraphics[scale=0.5]{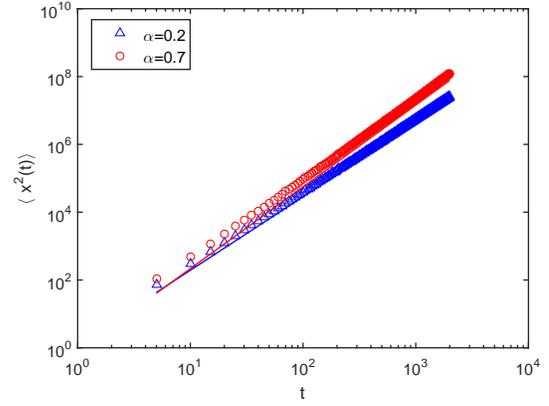}
\caption{MSD of the stochastic process described by the Langevin picture of the L\'{e}vy-walk-type model with strong correlated waiting times in the case of friction coefficient $\gamma=0$.  Solid lines represent the theoretical values (\ref{MSD0}) and the markers are simulation results. The parameters are, respectively, taken as $\gamma=0, D_v=1$, $\beta=1$, and the initial velocity is $v_0=0$. }\label{MSD_gama_0}
\end{figure}

In this subsection, the Langevin system without friction is considered
\begin{equation}\label{model2}
\begin{split}
\frac{dx(t)}{dt}&=v(t),\\
\frac{dv(s)}{ds}&=\eta(s),\\
\frac{dt(s)}{ds}&=\int_0^sds'M(s-s')\tau(s').
\end{split}
\end{equation}
Here we also assume the strong correlation of waiting times $M(s)=1$.
The correlation function of the process $v(s)$ could be easily obtained by the correlation function $\langle \eta(s_1)\eta(s_2)\rangle=2D_v\delta(s_1-s_2)$ of white Gaussian noise,
\begin{equation}\label{B}
\begin{split}
\langle v(s_1)v(s_2)\rangle&=\int_0^{s_1}ds'_1\int_0^{s_2}ds'_2\langle\eta(s'_1)\eta(s'_2)\rangle\\
&=2D_v \textrm{min}\{s_1,s_2\}.
\end{split}
\end{equation}
Substituting (\ref{B}) and (\ref{h_s12}) into (\ref{vv}), and making some complicated calculations, the correlation function of $v(t)$ in Laplace space for small $\lambda_1$ and $\lambda_2$ is
\begin{equation}
\begin{split}
\langle v(\lambda_1)v(\lambda_2)\rangle\simeq K_6\frac{\lambda_1^{-\frac{1}{1+\alpha}}+\lambda_2^{-\frac{1}{1+\alpha}}}
{(\lambda_1+\lambda_2)^{1+2\tilde\alpha}},
\end{split}
\end{equation}
where $K_6=2^{\frac{1-\alpha}{1+\alpha}}D_v\Gamma(\tilde\alpha)
\Gamma(2/(1+\alpha))\alpha(1+\alpha)^{-\tilde\alpha}$. After making the inverse Laplace transform, the correlation function of the velocity process for long times is
\begin{equation}
\begin{split}
\langle v(t_1)v(t_2)\rangle
\simeq K_7(t_2-t_1)^{-\tilde\alpha}t_1^{2\tilde\alpha},
\end{split}
\end{equation}
where $K_7=K_6\frac{1}{\Gamma(1/(1+\alpha))}\frac{1}{\Gamma(1+2\tilde\alpha)}$ and $t_2>t_1$.
Finally, one arrives at the correlation function of stochastic process $x(t)$
\begin{equation}
\begin{split}
\langle & x(t_1)x(t_2)\rangle \\
&\simeq\Theta(t_2-t_1)K_7\Bigg[\frac{1}{2+\tilde\alpha}
B\left(1+2\tilde\alpha,\frac{1}{1+\alpha}\right)t_1^{2+\tilde\alpha}\\
&~~~+(1+\alpha)B\left(1+2\tilde\alpha,\frac{1}{1+\alpha}+1;\frac{t_1}{t_2}\right)t_2^{2+\tilde\alpha}\Bigg]\\
&~~~+\Theta(t_1-t_2)K_7\Bigg[\frac{1}{2+\tilde\alpha}
B\left(1+2\tilde\alpha,\frac{1}{1+\alpha}\right)t_2^{2+\tilde\alpha}\\
&~~~+(1+\alpha)B\left(1+2\tilde\alpha,\frac{1}{1+\alpha}+1;\frac{t_2}{t_1}\right)t_1^{2+\tilde\alpha}\Bigg].
\end{split}
\end{equation}
When $t_1=t_2$, the MSD of $x(t)$ for long times is
\begin{equation}\label{MSD0}
\langle x^2(t)\rangle
\simeq K_7\frac{2B\left(1+2\tilde\alpha,\frac{1}{1+\alpha}\right)}{2+\tilde\alpha} t^{2+\tilde\alpha},
\end{equation}
which is a hyperdiffusion phenomenon. The consistency of the simulation results and analytical ones of the MSD for different $\alpha$ is shown in Fig. \ref{MSD_gama_0}. Contrary to the case $\gamma\neq0$, with the increase of $\alpha$, the diffusion becomes faster in the non-friction case.
The correlated power-law distributed waiting times suppresses the diffusion of velocity $v$ and thus $x$ due to the occasionally long waiting
time, which implies the diffusion with smaller $\alpha$ is more suppressed.

Another way to obtain the moments of the stochastic process $x(t)$ described by model (\ref{model2}) with $\beta=1$ is utilizing the self-similar property. In the case of $\gamma=0$ and $t>0$, (\ref{self}) could be rewritten as
\begin{equation}
x(t)\overset{d}{=}t^{1+\tilde\alpha/2}x(1).
\end{equation}
Therefore, the moments of stochastic process $x(t)$ is
\begin{equation}\label{moments}
\left\langle x^{2n}(t)\right\rangle=t^{n(2+\tilde\alpha)}\left\langle x^{2n}(1)\right\rangle.
\end{equation}
From \cite{FriedrichJenkoBauleEulePRE:2006}, the moments of the stochastic process described by L\'{e}vy-walk-type model with uncorrelated heavy-tailed waiting times for $\gamma=0$ are
\begin{equation}
\left\langle x^{2n}(t)\right\rangle\propto t^{n(2+\alpha)}.
\end{equation}
Comparing these two models, i.e., L\'{e}vy-walk-type model with strong correlated waiting times (\ref{model2}) and uncorrelated waiting times in \cite{FriedrichJenkoBauleEulePRE:2006}, whose moments are $t^{n(2+\tilde\alpha)}$ and $t^{n(2+\alpha)}$ respectively, one can observe the slower diffusion phenomenon because of the correlation of waiting times.

Similarly to the case $\gamma\neq0$, for fixed $t_1$ and $t_2\rightarrow \infty$, the correlation coefficient $r[x(t_1),x(t_2)]$ of the stochastic process $x(t)$ for $\gamma=0$ can be obtained
\begin{equation}\label{co2}
\begin{split}
r[x(t_1),x(t_2)]=\frac{\langle x(t_1)x(t_2)\rangle}{\sqrt{\langle x^2(t_1)\rangle\langle x^2(t_2)\rangle}}
\simeq K_8 \left(\frac{t_1}{t_2}\right)^{3\tilde\alpha/2},
\end{split}
\end{equation}
where $K_8=\frac{2+5\alpha+3\alpha^2}{2B\left(1+2\tilde\alpha,\frac{1}{1+\alpha}\right)(1+3\alpha)}$. That is to say, the stochastic process $x(t)$ described by model (\ref{model2}) without the effect of friction is also long-range dependent since $0<3\tilde\alpha/2<1$. Comparing the correlation coefficient (\ref{co1}) and (\ref{co2}), the stronger correlation of the stochastic process is observed in the case of friction coefficient $\gamma\neq0$ than the non-friction case.

For the uncorrelated waiting times case, i.e., $dt(s)/ds=\tau(s)$ in model (\ref{model2}), after some similar calculations, one can obtain the correlation function ($t_2>t_1$) and MSD of the stochastic process $x(t)$
\begin{equation*}
\begin{split}
\langle x(t_1)x(t_2)\rangle &=\frac{2D_v}{\Gamma(2+\alpha)}t_2t_1^{1+\alpha}-\frac{2D_v\alpha}{\Gamma(3+\alpha)}t_1^{2+\alpha},\\
\langle x^2(t)\rangle &=\frac{4D_v}{\Gamma(3+\alpha)}t^{2+\alpha}.
\end{split}
\end{equation*}
Then the correlation coefficient is 
\begin{equation}\label{co3}
\begin{split}
r_0[x(t_1),x(t_2)]=\frac{\langle x(t_1)x(t_2)\rangle}{\sqrt{\langle x^2(t_1)\rangle\langle x^2(t_2)\rangle}}
\simeq \frac{2+\alpha}{2} \left(\frac{t_1}{t_2}\right)^{\alpha/2},
\end{split}
\end{equation}
for fixed $t_1$ and $t_2\rightarrow \infty$. In the case of friction coefficient $\gamma=0$, the weaker correlation of the stochastic process $x(t)$ is detected in the L\'{e}vy-walk-type model with correlated waiting times in model (\ref{model2}) since $3\tilde{\alpha}/2>\alpha/2$, being contrary to the case $\gamma\neq0$, where the stronger correlation of the stochastic process $x(t)$ is led to when the waiting times are correlated.

\subsection{Correlated waiting times with $0<\beta<1$ in the cases $\gamma=0$ and $\gamma\neq0$}\label{IIIC}
\begin{figure}
\begin{minipage}{0.35\linewidth}
  \centerline{\includegraphics[scale=0.3]{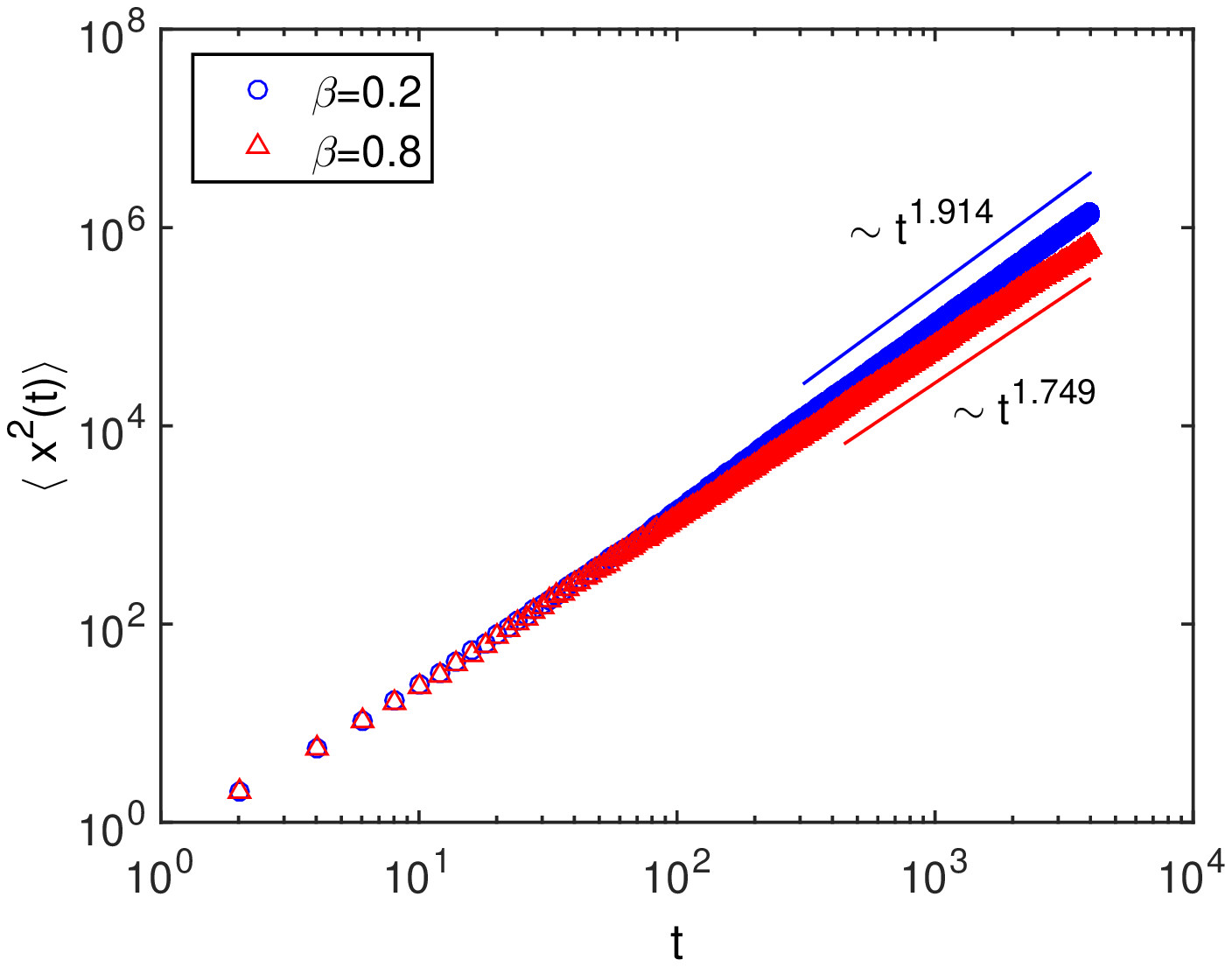}}
  \centerline{(a)}
\end{minipage}
\hspace{1cm}
\begin{minipage}{0.35\linewidth}
  \centerline{\includegraphics[scale=0.3]{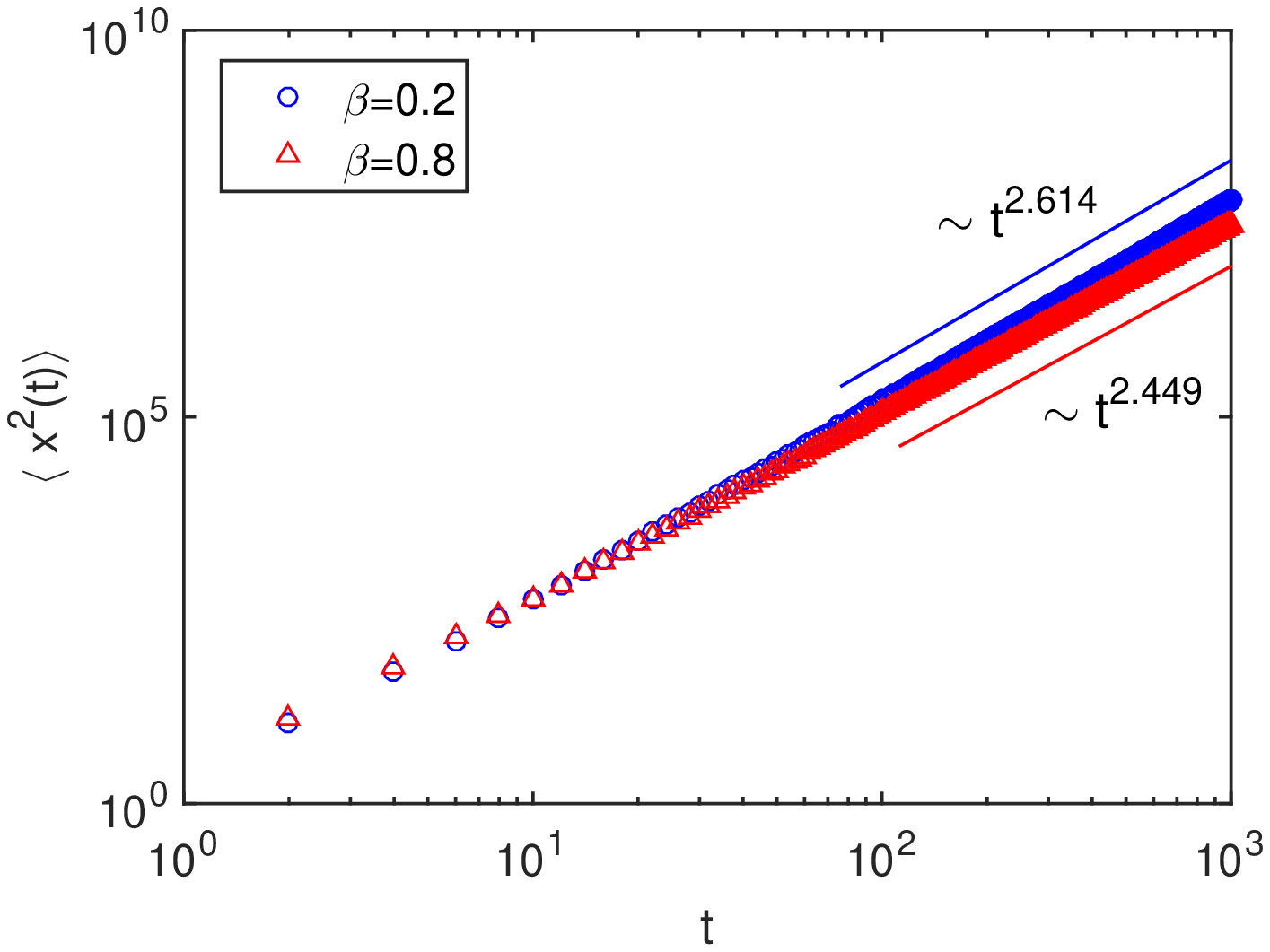}}
  \centerline{(b)}
\end{minipage}
\caption{Simulation results of the MSD of stochastic process described by the L\'{e}vy-walk-type model with correlated waiting times for different $\gamma$ and $\beta$. The parameters are respectively taken as $D_v=1$, $\alpha=0.7$, $\gamma=2$ (left panel) or $\gamma=0$ (right panel), and the initial velocity $v_0=0$. Blue circle-markers and red triangle-markers represent the cases of $\beta=0.2$ and $\beta=0.8$, respectively.
The solid lines in (a) and (b) denote the theoretical slopes in \eqref{x} and \eqref{gama0}, respectively.
Big $\alpha$ could increase the differences between the slopes of the two cases ($\beta=0.2$ and $\beta=0.8$). Hence we choose $\alpha=0.7$ to distinguish the two simulation results,  evidently.
}\label{beta}
\end{figure}

In this subsection, we consider the case that the memory function $M(s)=s^{\beta-1}/\Gamma(\beta)$ with $0<\beta<1$.
Compared to the situation $M(s)\equiv1$, $M(s)=s^{\beta-1}/\Gamma(\beta)$ with $0<\beta<1$ is a monotone decreasing function, which implies that with the increment of the time difference, the impact of historical waiting times on the current waiting time becomes weaker. On the other hand, with the increase of $\beta$, this impact becomes larger, until it reaches the maximum value when $\beta=1$.
We still focus on the MSD $\langle x^2(t)\rangle$ to analyze the diffusion phenomenon. The calculation is similar to the above situation ($\beta=1$), so we will not show it in detail here. In the effect of friction, $\gamma\neq0$, inserting the function $\mu(s)=s^\beta/\Gamma(1+\beta)$ into (\ref{h}), and using (\ref{vv}) and (\ref{vsvs}), after some complicated calculations, the correlation function of velocity process $v(t)$ in Laplace space for small $\lambda_1$ and $\lambda_2$ is
\begin{equation}
\begin{split}
\langle v(\lambda_1)v(\lambda_2)\rangle\propto \frac{\lambda_1^{-1}\lambda_2^{\alpha-1}+\lambda_2^{-1}\lambda_1^{\alpha-1}}
{(\lambda_1+\lambda_2)^{\alpha/(\alpha\beta+1)}}.
\end{split}
\end{equation}
As for the double integral of $\langle v(t_1)v(t_2)\rangle$, the correlation function of the stochastic process $x(t)$ in Laplace space is
\begin{equation}
\begin{split}
\langle x(\lambda_1)x(\lambda_2)\rangle&=\frac{1}{\lambda_1\lambda_2}\langle v(\lambda_1)v(\lambda_2)\rangle \\ &\propto \frac{1}{\lambda_1\lambda_2}\frac{\lambda_1^{-1}\lambda_2^{\alpha-1}+\lambda_2^{-1}\lambda_1^{\alpha-1}}
{(\lambda_1+\lambda_2)^{\alpha/(\alpha\beta+1)}}.
\end{split}
\end{equation}
Then perform the inverse Laplace transform and take $t_1=t_2$, 
which leads to the MSD of the stochastic process $x(t)$ as
\begin{equation}\label{x}
\begin{split}
\langle x^2(t)\rangle
\propto t^{2-\frac{\alpha^2\beta}{\alpha\beta+1}},
\end{split}
\end{equation}
displaying a sub-ballistic superdiffusion phenomenon.
When $\beta=0$, (\ref{x}) becomes $\langle x^2(t)\rangle\simeq t^2$, showing the same dynamical behavior as the L\'{e}vy walk model with independent heavy-tailed waiting times. For $\beta=1$, (\ref{x}) arrives at $\langle x^2(t)\rangle\simeq t^{2-\alpha\tilde\alpha}$, which coincides with the case of strong correlated waiting times (\ref{MSD}).
From (\ref{x}), we observe the phenomenon that with the increase of $\beta$, i.e., the correlation of waiting times becomes stronger, the diffusion of the particles slows down. That could also be seen in Fig. \ref{beta}(a), which shows the simulation results of the MSD (\ref{x}) for different $\beta$.

For the case of friction coefficient $\gamma=0$, with the similar calculation procedures, we obtain the MSD of the stochastic process $x(t)$
\begin{equation}\label{gama0}
\langle x^2(t)\rangle
\propto t^{2+\frac{\alpha}{\alpha\beta+1}},
\end{equation}
which is a super-ballistic diffusion phenomenon.
For $\beta=0$, (\ref{gama0}) reduces to the results given in \cite{FriedrichJenkoBauleEulePRE:2006}. While for $\beta=1$, (\ref{gama0}) coincides with the strong correlated waiting times case (\ref{MSD0}). Like the above situation $\gamma\neq0$, here with the increase of $\beta$, the diffusion also slows down. The simulation results of the MSD (\ref{gama0}) for different $\beta$ are shown in Fig. \ref{beta}(b).

\section{Conclusion}\label{four}
Since the CTRW was put forward by Montroll and Weiss in 1965, it has been widely concerned and developed because of its practicability and flexibility. 
Recently, the correlated CTRWs, which could describe the stochastic processes with memory effect, are further developed and discussed. As an important and sometimes fundamental supplements to CTRWs, Langevin pictures attract a lot of researchers' interest and sometime play an irreplaceable role. 
In this paper, we pay attention to the Langevin picture of the L\'{e}vy-walk-type model with correlated waiting times, which is an aging and nonstationary process. We find that it displays a sub-ballistic superdiffusion in the effect of friction and hyperdiffusion without friction, which are both slower than the diffusion of the L\'{e}vy-walk-type model with uncorrelated waiting times. The stronger of the correlation of waiting times is, the slower the diffusion becomes whether there is friction or not. Besides that, the long-range dependence and scaling property are discovered in our model. The weak ergodicity breaking is also detected, which is similar to the one of the  L\'{e}vy-walk-type model with uncorrelated waiting times.

\section*{Acknowledgments}
This work was supported by the National Natural Science Foundation of China under grant no. 11671182, and the Fundamental Research Funds for the Central Universities under grants no. lzujbky-2018-ot03 and no. lzujbky-2017-ot10.

\appendix
\section{Derivation of \eqref{ADD3}}\label{App1}
Since the derivation processes of the two factors in \eqref{ADD3} are similar, we just present one of them, e.g.,
\begin{equation}
  \begin{split}
    I:&=\left\langle \exp\left\{-\lambda_2\int_{s_1}^{s_2}ds'\tau(s')\mu(s_2-s')\right\}\right\rangle \\
&=\exp\left\{-\int_0^{s_2-s_1}ds'[\lambda_2\mu(s_2-s_1-s')]^\alpha\right\}.
  \end{split}
\end{equation}
We will make best use of the properties of the $\alpha$-stable totally skewed L\'{e}vy process $t_0(s)=\int_0^s ds'\tau(s')$, i.e., the increment is stationary: 
\begin{equation}\label{stationary}
  t_0(s_2)-t_0(s_1)\overset{d}{=}t_0(s_2-s_1),
\end{equation}
and independent. We first divide the integral interval $(s_1,s_2)$ into $N$ parts with the same length $\Delta s$ ($N\Delta s=s_2-s_1$). Then we approximate the integral in $I$ by It\^{o} interpretation \cite{Ito:1950,Ito:1944} (the result will be the same for other interpretations, such as Stratonovich \cite{Stratonovich:1966} or H\"{a}nggi-Klimontovich \cite{Hanggi:1982}):
\begin{equation}\label{ADD4}
  I=\lim_{N\rightarrow\infty} \left\langle \exp\left\{-\lambda_2 \sum_{i=1}^N \mu(s_2-s'_i)[t_0(s'_i+\Delta s)-t_0(s'_i)]\right\}\right\rangle.
\end{equation}
Considering the independence of the increments, the ensemble average in \eqref{ADD4} can be divided into $N$ parts for $N$ independent increments $t_0(s'_i+\Delta s)-t_0(s'_i)$.
Using the stationarity of the increment \eqref{stationary}, together with the characteristic function of $t_0(s)$ \eqref{character}, the integral $I$ can be reduced to
\begin{equation}
\begin{split}
    I&=\lim_{N\rightarrow\infty} \prod_{i=1}^N \exp\left\{-\Delta s [\lambda_2\mu(s_2-s'_i)]^\alpha\right\}  \\
    &=\lim_{N\rightarrow\infty} \exp\left\{-\sum_{i=1}^N \Delta s [\lambda_2\mu(s_2-s'_i)]^\alpha\right\}  \\
    &=\exp\left\{-\int_{s_1}^{s_2}ds' [\lambda_2\mu(s_2-s')]^\alpha\right\}  \\
    &=\exp\left\{-\int_0^{s_2-s_1}ds'[\lambda_2\mu(s_2-s_1-s')]^\alpha\right\}.
\end{split}
\end{equation}

\section*{References}
\bibliographystyle{apsrev4-1}
\bibliography{Reference}

\end{document}